\documentclass[amsmath,reprint,prb,longbibliography,superscriptaddress,bibnotes]{revtex4-2}
\usepackage{amsthm,amssymb,amsfonts,graphicx,verbatim, xcolor,bm} 
\usepackage{hyperref} 
\usepackage[utf8]{inputenc} 
\usepackage{siunitx} 
\usepackage{braket}
\usepackage{ulem}

\begin{document}

\title{Superfluid stiffness bounds in time-reversal symmetric superconductors}
\author{Yongxin Zeng}
\email{yz4788@columbia.edu}
\affiliation{Department of Physics, Columbia University, New York, NY 10027}
\author{Andrew J. Millis}
\affiliation{Department of Physics, Columbia University, New York, NY 10027}
\affiliation{Center for Computational Quantum Physics, Flatiron Institute, New York, NY 10010}

\begin{abstract}
Quantum geometry has been shown to make an important contribution to the superfluid stiffness of superconductors, especially for flat-band systems such as moir\'e materials. In this work we use mean-field theory to derive an expression for the superfluid stiffness of time-reversal symmetric superconductors at zero temperature by computing the energy of the mean-field ground state as a function of pairing momentum. We show that the quantum geometric contribution to superfluid stiffness is a consequence of broken Galilean invariance in the interaction Hamiltonian, arising from momentum-dependent form factors related to the momentum dependence of Bloch states. The effects of broken Galilean invariance are not fully parametrized by the quantum metric considered in previous work. We obtain general lower and upper bounds that apply to both continuum and lattice models and present numerical calculations of the precise value in several important cases. The superfluid stiffness of superconductivity in a Landau level saturates the lower bound and the superfluid stiffness of the other cases we consider is close to the general lower bound we derive.  In multilayer rhombohedral graphene the geometric contribution is shown not to be the dominant contribution to the superfluid stiffness, despite the flat band behavior in the vicinity of the Fermi level. Finally, assuming contact interaction and uniform pairing, we show that the superfluid stiffness is proportional to the ``minimal quantum metric" introduced in previous work. We provide a continuum version of the minimal quantum metric and explain its physical origin.
\end{abstract}

\maketitle

{\it Introduction.---}
The superfluid stiffness is an important characteristic of a superconductor that measures the energy cost of a spatially varying U(1)-phase of the superconducting order parameter; it parametrizes the maximal value of the supercurrent and in two-dimensional materials sets the value of the Berezinskii-Kosterlitz-Thouless (BKT) temperature below which superconductivity can appear.  In the conventional Bardeen-Cooper-Schrieffer (BCS) theory of superconductivity \cite{bardeen1957theory}, the superfluid stiffness is written as $D = n_s/m_{\rm eff}$ where $n_s$ is the density of paired electrons and $m_{\rm eff}$ is the electron effective mass. Recent studies of superconductivity and exciton condensation in flat-band systems such as moir\'e materials \cite{cao2018unconventional, yankowitz2019tuning, park2021tunable, tian2023evidence, banerjee2025superfluid, tanaka2025superfluid} have revealed another contribution to superfluid stiffness that arises from the nontrivial quantum geometry of the electronic Bloch states \cite{peotta2015superfluidity, julku2016geometric, liang2017band, torma2022superconductivity, xie2020topology, hu2019geometric, julku2020superfluid, chen2024ginzburg, huhtinen2022revisiting, peotta2025quantum, tam2024geometry, jiang2024geometric, herzog2022many, herzog2022superfluid, mao2023diamagnetic, mao2024upper, hazra2019bounds, verma2021optical, hu2022exciton, verma2024exciton, han2024quantum, lamponen2025superconductivity, torma2023essay, yu2024quantum}. This quantum geometric contribution to superfluid stiffness makes possible a current-carrying state and a non-zero BKT transition temperature even in ideal flat-band situations where the carrier mass $m_{\rm eff}\to\infty$.

While the quantum geometric contribution to superfluid stiffness is well-defined and can be evaluated numerically in the general case \cite{peotta2015superfluidity}, simple analytic results are obtained only for special lattice models with attractive onsite Hubbard interactions and superconducting states that satisfy the ``uniform pairing conditions" described below (see Ref.~\cite{han2024quantum} for generalization to a wider class of ideal models). In this situation, it was argued that the quantum geometric contribution to the superfluid stiffness is determined by the momentum-space integral of the quantum metric of the electronic Bloch states \cite{peotta2015superfluidity, julku2016geometric, liang2017band, chen2024ginzburg}. It was subsequently noticed \cite{simon2020contrasting} that the quantum metric depends on the sublattice embedding while the superfluid stiffness does not, and it was shown that the difficulty could be resolved if the integral of quantum metric is replaced by its {\it minimal} value for all possible sublattice embeddings \cite{huhtinen2022revisiting, herzog2022many, peotta2025quantum, tam2024geometry, jiang2024geometric}. Relaxing the restrictive assumptions of onsite interactions and the uniform pairing conditions in order to understand the quantum geometric contribution to superfluid stiffness under more general conditions is an important goal addressed in this paper.

The superfluid stiffness of a superconductor can be computed equivalently by linear response theory and by taking derivatives of the thermodynamic potential with respect to the pairing momentum \cite{scalapino1993insulator}. The former approach involves the coupling of the superconducting order parameter with the vector potential and thus requires careful attention to gauge invariance and to virtual interband processes \cite{huhtinen2022revisiting, mao2023diamagnetic, mao2024upper}. In this work we adopt the latter approach as it allows single-band projection of electron-electron interactions and avoids explicit consideration of interband tunneling processes. We show that the essence of the quantum geometric contribution to superfluid stiffness is the breaking of Galilean invariance by the interaction terms in the projected Hamiltonian, arising from form factors with a structure inherited from quantum geometry-induced momentum dependence of Bloch wave functions. The form factors involve inner products of Bloch wave functions at different momenta, which in the general case cannot be simplified to local geometric quantities such as the quantum metric or the related ``minimal quantum metric". Within mean-field theory, we provide lower and upper bounds on superfluid stiffness; the bounds apply to both lattice and continuum models with attractive electron-electron interactions. We show that the lower bound saturates for the ideal quantum geometry of Landau levels and closely approximates the exact mean-field superfluid stiffness in the examples we present. In the special case of contact interactions and uniform pairing conditions, we reproduce the expression in terms of the minimal quantum metric proposed in previous work and generalize it to continuum models.

{\it General formalism.---}
Consider a multi-band system in which electrons interact via an attractive potential. The single-electron states are labeled by a band index, a Bloch momentum $\bm k$, and an internal index $\sigma=\uparrow,\downarrow$ for a pair of time-reversal partners (e.g., spin or valley). We assume that only one band (per spin) is active, i.e., all other bands are either empty or fully filled and are separated from the active band by a large energy gap. The active band is described by its energy dispersion $\epsilon_{\bm k \uparrow} = \epsilon_{-\bm k \downarrow} \equiv \epsilon_{\bm k}$ and Bloch eigenstates $\ket{u_{\bm k \uparrow}} = \ket{u_{-\bm k \downarrow}}^* \equiv \ket{u_{\bm k}}$. Projecting onto the active band, the Hamiltonian $H = H_0 + H_{\rm int}$ consists of a single-particle term $H_0 = \sum_{\bm k \sigma} \epsilon_{\bm k \sigma} c_{\bm k \sigma}^{\dagger} c_{\bm k \sigma}$ and an interaction term
\begin{equation}
    H_{\rm int} = -\frac 1A \sum_{\bm k \bm k' \bm q} V_{\bm{kk}'}(\bm q) c_{\bm q+\bm k \uparrow}^{\dagger} c_{\bm q-\bm k \downarrow}^{\dagger} c_{\bm q-\bm k' \downarrow} c_{\bm q+\bm k' \uparrow}.
\end{equation}
Here $A$ is the size (area or volume) of the system and $V_{\bm{kk}'}(\bm q)$ is the band-projected electron-electron interaction potential that encodes information about quantum geometry of the active band; its explicit expression in terms of form factors will be presented in the examples below. Because of time-reversal symmetry, $V_{\bm{kk}'}(-\bm q) = V_{\bm{kk}'}^*(\bm q)$. In particular, $V_{\bm{kk}'}(\bm q=0) \in \mathbb{R}$ and $\partial_{\bm q} V_{\bm{kk}'}(\bm q) |_{\bm q=0}$ is purely imaginary.

Assuming spin-singlet pairing, a general BCS state with pairing momentum $\bm q$ takes the form
\begin{equation}
    \ket{\Psi_{\bm q}} = \prod_{\bm k} \left[\cos\frac{\theta_{\bm k}(\bm q)}{2} + e^{i\phi_{\bm k}(\bm q)} \sin\frac{\theta_{\bm k}(\bm q)}{2} c_{\bm q+\bm k \uparrow}^{\dagger} c_{\bm q-\bm k \downarrow}^{\dagger} \right] \ket{0},
\end{equation}
where $\ket{0}$ is the vacuum state in which the active band is empty. $\theta_{\bm k}(\bm q) \in [0,\pi]$ and $\phi_{\bm k}(\bm q) \in [0,2\pi)$ are variational parameters that are chosen to minimize the total energy of the BCS state $E_{\bm q}[\theta,\phi] \equiv \braket{\Psi_{\bm q} | H | \Psi_{\bm q}} = E_{\bm q}^0 + E_{\bm q}^{\rm int}$ at fixed total electron number $N_e = 2\sum_{\bm k} \sin^2 (\theta_{\bm k}/2)$, or equivalently, to minimize $E_{\bm q}-\mu N_e$ at fixed chemical potential $\mu$ as detailed in the Supplemental Material (SM) \cite{SM}. The total momentum of the state $\bm P$ is related to the pairing momentum $\bm q$ by $\bm P = N_e \bm q$.

Explicitly, the single-particle energy is
\begin{equation}
    E_{\bm q}^0[\theta] = \sum_{\bm k} (\epsilon_{\bm k + \bm q} + \epsilon_{\bm k - \bm q}) \sin^2 \frac{\theta_{\bm k}(\bm q)}{2},
\end{equation}
and interaction energy is
\begin{align}
    E_{\bm q}^{\rm int}[\theta,\phi] = - \frac{1}{4A} \sum_{\bm{kk}'} &V_{\bm{kk}'}(\bm q) e^{i[\phi_{\bm k'}(\bm q) - \phi_{\bm k}(\bm q)]} \nonumber\\
    &\times \sin \theta_{\bm k}(\bm q) \sin \theta_{\bm k'}(\bm q).
\end{align}
 The superfluid stiffness at zero temperature is obtained by taking the second derivative of the total energy $E_{\bm q}$ with respect to the pairing momentum $\bm q$:
\begin{equation}
    D_{ij} = \frac{1}{A} \frac{d^2 E_{\bm q}}{dq_i dq_j}\Bigg|_{\bm q=0}.
\end{equation}
As in Ref.~\cite{huhtinen2022revisiting}, we denote by $d/dq_i$ the total derivative with respect to $q_i$ that takes into account the change of $(\theta_{\bm k}, \phi_{\bm k})$ with $\bm q$ as well as the dependence of $\varepsilon_{\bm k+\bm q}-\varepsilon_{\bm k-\bm q}$ and of $V_{\bm{kk}'}(\bm q)$ on $\bm q$. The precise determination of the BCS ground state and its energy in general requires self-consistent calculations at each $\bm q$. The problem is greatly simplified for systems with Galilean invariance, i.e., the boosted Hamiltonian $H - \bm V \cdot \bm P$ ($\bm V$: velocity of the moving frame; $\bm P$: total momentum operator) is equivalent to $H$ up to constant momentum and energy shifts as well as gauge transformations,
which guarantees that $\theta_{\bm k}(\bm q)$ and $E_{\bm q}^{\rm int}$ are independent of $\bm q$. The superfluid stiffness then takes the conventional form
\begin{equation} \label{eq:Ds_Gal_inv}
    D_{ij} = \frac{2}{A} \sum_{\bm k} \frac{\partial^2 \epsilon_{\bm k}}{\partial k_i \partial k_j} \sin^2\frac{\theta_{\bm k}}{2} = \frac{n_e}{m_{\rm eff}} \delta_{ij},
\end{equation}
where $n_e=N_e/A$ is the total electron density and $m_{\rm eff} = (\partial^2 \epsilon_{\bm k}/\partial k^2)^{-1}$ is the effective mass. Eq.~\eqref{eq:Ds_Gal_inv} holds regardless of the pairing symmetry and the precise form of interactions. The quantum geometric contribution to superfluid stiffness, to be discussed below, should therefore be viewed as a consequence of broken Galilean invariance due to the nontrivial momentum dependence of Bloch wave functions.

At $\bm q=0$, because $V_{\bm{kk}'}(0)\in\mathbb{R}$, the BCS ground state is real and its $\bm q$-derivative is purely imaginary: $\phi_{\bm k}(0) = 0$, $\partial_{\bm q} \theta_{\bm k}(\bm q)|_{\bm q=0} = 0$. As a result, the second derivative of $E_{\bm q}$ involves only the $\bm q$-dependence of $\phi_{\bm k}(\bm q)$ but not that of $\theta_{\bm k}(\bm q)$ (see SM \cite{SM}). Given $\theta_{\bm k} \equiv \theta_{\bm k}(\bm q=0)$, the mean-field superfluid stiffness can be formally expressed as
\begin{widetext}
\begin{equation} \label{eq:D_max_phi}
    D_{ij} = \frac{2}{A} \sum_{\bm k} \frac{\partial^2 \epsilon_{\bm k}}{\partial k_i \partial k_j} \sin^2\frac{\theta_{\bm k}}{2} -\frac{1}{4A^2} \frac{\partial^2}{\partial q_i \partial q_j} \max_{\phi} \left\{ \sum_{\bm{kk}'} \sin \theta_{\bm k} \sin \theta_{\bm k'}\, V_{\bm{kk}'}(\bm q) e^{i[\phi_{\bm k'}(\bm q) - \phi_{\bm k}(\bm q)]} \right\},
\end{equation}
\end{widetext}
where the first term on the right-hand side is the conventional contribution $D_{ij}^{\rm conv}$ due to band dispersion and the second term is the quantum geometric contribution $D_{ij}^{\rm geom}$.

A lower bound for the diagonal components ($i=j$) of $D^{\rm geom}$ is easily obtained by taking the absolute value of each term in the summation:
\begin{equation} \label{eq:Ds_lower}
    D_{ii}^{\rm geom} \ge -\frac{1}{4A^2} \sum_{\bm{kk}'} \frac{\partial^2}{\partial q_i^2} \big|V_{\bm{kk}'}(\bm q)\big| \sin \theta_{\bm k} \sin \theta_{\bm k'}.
\end{equation}
The lower bound saturates if the phase of $V_{\bm{kk}'}(\bm q)$ can be written as $\arg [V_{\bm{kk}'}(\bm q)] = \varphi_{\bm k}(\bm q) - \varphi_{\bm k'}(\bm q)$ so that the energy is minimized with $\phi_{\bm k}(\bm q) = \varphi_{\bm k}(\bm q)$ for all $\bm q$. A prominent example is the lowest Landau level most conveniently formulated in the Landau gauge as shown in SM \cite{SM}. In the general case, however, any given choice of $\phi_{\bm k}(\bm q)$ provides an upper bound for $E_{\bm q}$ and therefore an upper bound for $D$. One possible choice is $\phi_{\bm k}(\bm q) = \arg\braket{u_{\bm k+\bm q} | u_{\bm k-\bm q}}$ and it follows that
\begin{align} \label{eq:Ds_upper}
    D_{ii}^{\rm geom} \le -\frac{1}{4A^2} \sum_{\bm{kk}'} \frac{\partial^2}{\partial q_i^2} &\left[|V_{\bm{kk}'}(\bm q)| e^{i\varphi_{\bm{kk}' \bm q}}\right]  \sin \theta_{\bm k} \sin \theta_{\bm k'}
\end{align}
with
\begin{equation} \label{eq:phi_kkq}
    \varphi_{\bm{kk}'\bm q} \equiv \arg\left[V_{\bm{kk}'}(\bm q) \braket{u_{\bm k-\bm q} | u_{\bm k+\bm q}} \braket{u_{\bm k'+\bm q} | u_{\bm k'-\bm q}}\right].
\end{equation}
The advantage of this choice of $\phi_{\bm k}(\bm q)$ is that $\varphi_{\bm{kk}'\bm q}$ is gauge invariant for band-projected density-density interaction $V_{\bm{kk}'}(\bm q)$. Below we provide examples including continuum and periodic-lattice systems with explicit expressions of $V_{\bm{kk}'}(\bm q)$ to illustrate the above formalism and compare the bounds of superfluid stiffness with its precise value within mean-field theory.

{\it Continuum band.---}
For a system with continuous translation symmetry such as the low-energy effective model of untwisted graphene multilayers, the low-energy physics lives in an unbounded momentum space and the projected interaction takes the form
\begin{equation} \label{eq:V_proj_cont}
    V_{\bm{kk}'}(\bm q) = V_{\bm k - \bm k'} \Lambda_{\bm k + \bm q, \bm k' + \bm q} \Lambda_{\bm k' - \bm q, \bm k - \bm q},
\end{equation}
where $V_{\bm k - \bm k'}$ is the unprojected interaction potential in momentum space and $\Lambda_{\bm{kk}'} \equiv \braket{u_{\bm k} | u_{\bm k'}}$ is the form factor. A lower bound for superfluid stiffness is readily obtained:
\begin{align} \label{eq:Ds_lower_cont}
    D_{ii}^{\rm geom} \ge - \frac{1}{4A^2} \sum_{\bm{kk}'} &V_{\bm k - \bm k'} \sin \theta_{\bm k} \sin \theta_{\bm k'} \nonumber\\ 
    &\times \frac{\partial^2}{\partial q_i^2} \left|\Lambda_{\bm k + \bm q, \bm k' + \bm q} \Lambda_{\bm k' - \bm q, \bm k - \bm q}\right|.
\end{align}
The upper bound that follows from Eq.~\eqref{eq:Ds_upper}
\begin{align} \label{eq:Ds_upper_cont}
    D_{ii}^{\rm geom} \le - \frac{1}{4A^2} &\sum_{\bm{kk}'} V_{\bm k - \bm k'} \sin \theta_{\bm k} \sin \theta_{\bm k'} \nonumber\\ 
    &\times \frac{\partial^2}{\partial q_i^2} \left(|\Lambda_{\bm k + \bm q, \bm k' + \bm q} \Lambda_{\bm k' - \bm q, \bm k - \bm q}| e^{i\varphi_{\bm{kk}'\bm q}} \right)
\end{align}
is gauge invariant since the phase
\begin{align}
    \varphi_{\bm{kk}'\bm q} \equiv \arg&[\braket{u_{\bm k-\bm q} | u_{\bm k+\bm q}} \braket{u_{\bm k+\bm q} | u_{\bm k'+\bm q}} \nonumber\\
    &\times \braket{u_{\bm k'+\bm q} | u_{\bm k'-\bm q}} \braket{u_{\bm k'-\bm q} | u_{\bm k-\bm q}}]
\end{align}
is the gauge-invariant discrete Berry phase around the closed loop $\bm k-\bm q \to \bm k+\bm q \to \bm k'+\bm q \to \bm k'-\bm q \to \bm k-\bm q$.

As an analytically solvable example, consider the ``ideal parent band" model in Ref.~\cite{tan2024parent} with form factor
\begin{equation} \label{eq:Landau_form_factor}
    \Lambda_{\bm{kk}'} = \exp\left\{-\frac{\mathcal{B}}{4} (|\bm k - \bm k'|^2 + 2i\bm k \times \bm k') \right\}
\end{equation}
such that the Berry curvature $\mathcal{B}$ and quantum metric $g_{ij} = \delta_{ij} \mathcal{B}/2$ are uniform in $\bm k$-space. Because
\begin{equation}
    \Lambda_{\bm k + \bm q, \bm k' + \bm q} = \Lambda_{\bm{kk}'} e^{i\mathcal{B}\bm q\times(\bm k - \bm k')/2},
\end{equation}
the projected interaction $V_{\bm{kk}'}(\bm q) = V_{\bm{kk}'}(0) e^{i\mathcal{B} \bm q \times (\bm k - \bm k')}$ and the lower bound saturates with $\phi_{\bm k}(\bm q) = \mathcal{B}\bm q \times \bm k$. However, since $|V_{\bm{kk}'}(\bm q)|$ is independent of $\bm q$, the geometric contribution $D^{\rm geom}$ vanishes. This is because the momentum-boosted form factor $\Lambda_{\bm k + \bm q, \bm k' + \bm q}$ is equivalent to $\Lambda_{\bm{kk}'}$ up to a gauge transformation. Despite the highly nontrivial quantum geometry of the Bloch states, Galilean invariance is effectively preserved and therefore quantum geometry does not contribute to superfluid stiffness. Another example of vanishing geometric superfluid stiffness is a spin-polarized Landau level as noted in Ref.~\cite{guerci2025fractionalization} and shown explicitly in SM \cite{SM} using the Landau gauge.

\begin{figure}
    \centering
    \includegraphics[width=\linewidth]{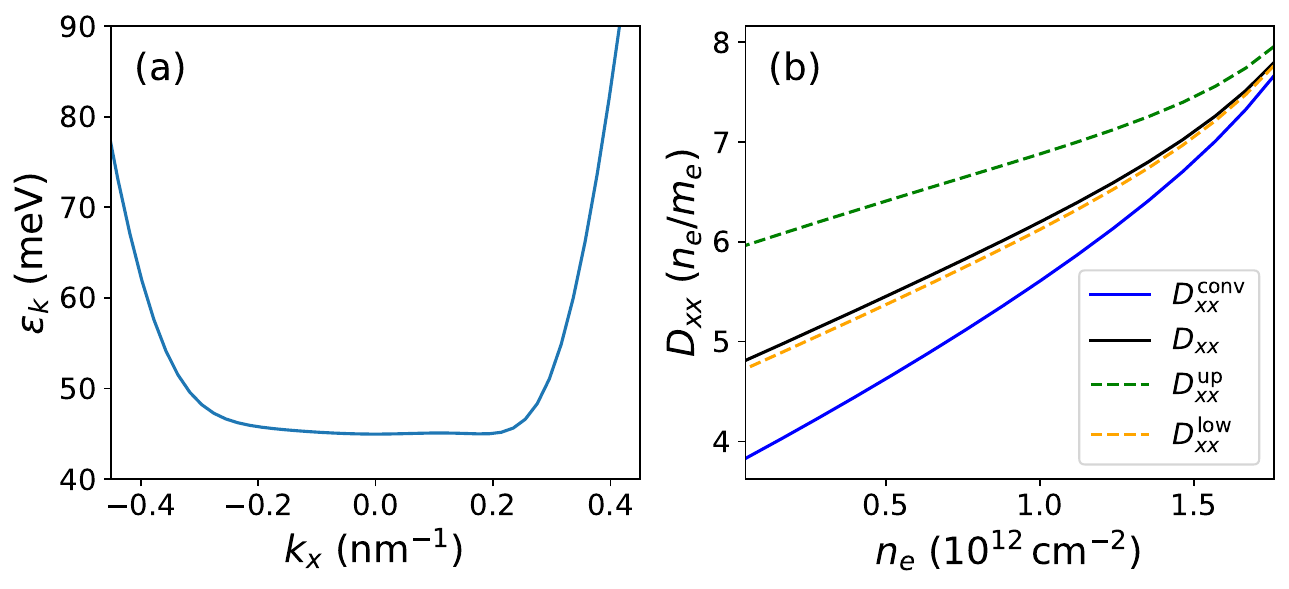}
    \caption{(a) The dispersion of the first conduction band of rhombohedral tetralayer graphene in a displacement field $u_D = \SI{30}{meV}$. Here the dispersion is plotted along the $k_x$-axis with the valley $K$-point set as the origin. (b) The $xx$-component of superfluid stiffness in units of $n_e/m_e$ ($n_e$: electron density; $m_e$: free electron mass). The blue and black solid lines represent the conventional contribution and the total $D_{xx}$ and the orange and green dashed lines represent the lower and upper bounds of the total $D_{xx}$ with geometric contributions from Eqs.~\eqref{eq:Ds_lower_cont} and \eqref{eq:Ds_upper_cont}.}
    \label{fig:graphene}
\end{figure}

As another example, consider rhombohedral tetralayer graphene in an out-of-plane displacement field that creates a potential difference $u_D = \SI{30}{meV}$ between neighboring layers. The displacement field opens a large gap between the first conduction and valence bands. When the system is lightly doped with electrons, only the first conduction band is active and the low-energy physics is described by a one-band Hamiltonian with dispersion shown in Fig.~\ref{fig:graphene}(a) and projected interaction in the form of Eq.~\eqref{eq:V_proj_cont}. Details of the microscopic model are provided in SM \cite{SM}.

As a model calculation, we assume contact interaction $V_{\bm k - \bm k'} = U = \SI{2}{eV.nm^2}$ and, assuming spin-polarized and valley-singlet pairing, perform self-consistent mean-field calculations at zero and finite pairing momentum $\bm q$; see Refs.~\cite{jiang2025quantum, kopnin2011surface} for related but slightly different calculations. From the mean-field ground state at $\bm q=0$ we calculate $D_{xx}^{\rm conv}$ as well as the lower and upper bounds of $D_{xx}^{\rm geom}$. The results in Fig.~\ref{fig:graphene}(b) show that $D_{xx}^{\rm geom}$ makes up a significant contribution to the total $D_{xx}$ at low carrier densities. We find that the exact mean-field superfluid stiffness from self-consistent calculations is well approximated by the lower bound while the upper bound is less tight. Despite the flatness of the band bottom, the sharp upturn at the edge of the flat-band region leads to a non-negligible $D_{xx}^{\rm conv}$ and therefore $D_{xx}^{\rm geom}$ is never the dominant contribution to the total $D_{xx}$.

{\it Periodic systems.---}
For periodic systems such as moir\'e materials, the momenta $\bm k,\bm k'$ live within a Brillouin zone with reciprocal lattice vectors $\bm G$. In continuum models, the projected interaction potential takes the form
\begin{align} \label{eq:V_proj_moire}
    V_{\bm{kk}'}(\bm q) = \sum_{\bm G} V_{\bm k-\bm k'+\bm G} &\braket{u_{\bm k+\bm q} | e^{i\bm G\cdot\bm r} | u_{\bm k'+\bm q}} \nonumber\\
    &\times\braket{u_{\bm k'-\bm q} | e^{-i\bm G\cdot\bm r} | u_{\bm k-\bm q}}.
\end{align}
A lower bound for $D_{ii}^{\rm geom}$ follows from Eq.~\eqref{eq:Ds_lower}. Similar to the continuum case, the upper bound that follows from Eq.~\eqref{eq:Ds_upper} is gauge invariant since $\varphi_{\bm{kk}'\bm q}$ (Eq.~\ref{eq:phi_kkq}) in this case involves a sum of closed-loop inner products that are individually gauge invariant.

\begin{figure}
    \centering
    \includegraphics[width=\linewidth]{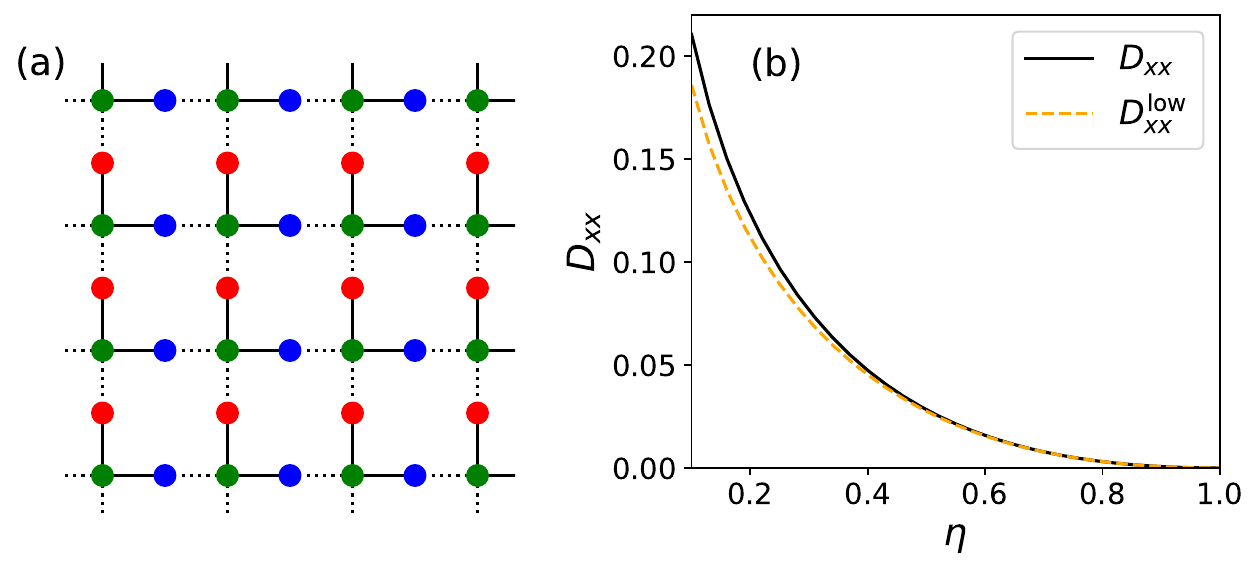}
    \caption{(a) Lieb lattice with staggered nearest-neighbor hopping. The solid and dotted lines represent hopping amplitude proportional to $1+\eta$ and $1-\eta$ respectively. (b) Superfluid stiffness of a half-filled Lieb lattice. The solid and dashed lines represent the exact mean-field superfluid stiffness from self-consistent calculations and the lower bound from Eq.~\eqref{eq:Ds_lower}.}
    \label{fig:lieb}
\end{figure}

Our formalism can also be applied to lattice models in which the Bloch states $\ket{u_{\bm k}} = \sum_{\alpha} u_{\bm k \alpha} \ket{\bm k \alpha}$ where $\ket{\bm k \alpha}$ is the plane-wave state on sublattice $\alpha$. As an example, we consider a Lieb lattice with staggered nearest-neighbor hopping as schematically illustrated in Fig.~\ref{fig:lieb}(a) and detailed in SM \cite{SM}. The single-particle band structure consists of a dispersionless middle band that is gapped from the upper and lower bands when the hopping asymmetry $\eta\ne 0$. We assume the gap is large enough such that interaction-induced band mixing is negligible. With onsite attractive interaction $U$, the projected interaction
\begin{equation} \label{eq:V_proj_Hubbard}
    V_{\bm{kk}'}(\bm q) = U\sum_{\alpha} u_{\bm k + \bm q, \alpha}^* u_{\bm k - \bm q, \alpha} u_{\bm k' - \bm q, \alpha}^* u_{\bm k' + \bm q, \alpha}
\end{equation}
is manifestly invariant under the embedding transformation $u_{\bm k \alpha} \to u_{\bm k \alpha} e^{i\bm k \cdot \bm{\tau}_{\alpha}}$ and therefore the lower bound \eqref{eq:Ds_lower} is also embedding-independent.

Now we assume the flat band is half-filled and compare the lower bound \eqref{eq:Ds_lower} with the mean-field superfluid stiffness from self-consistent calculations. Because the band is perfectly flat, the conventional contribution to superfluid stiffness vanishes and any nonzero superfluid stiffness is of geometric origin. Within the flat-band subspace, the interaction strength $U$ provides the only energy scale and we set $U=1$. The results in Fig.~\ref{fig:lieb}(b) show that the lower bound is very close to the exact value of mean-field superfluid stiffness and quantitatively reproduces the trend of $D_{xx}$ vanishing as $\eta\to 1$.

{\it Uniform pairing and minimal quantum metric.---}
The results presented so far show that quantum geometry enters superfluid stiffness as inner products of Bloch states (form factors) at different momenta that are in general not infinitesimally close to each other so that local geometric quantities such as the quantum metric are not directly relevant. To connect with previous work demonstrating a relation between the quantum metric and the geometric contribution to superfluid stiffness, we assume contact interaction $V_{\bm k-\bm k'} = U$ and the total interaction energy is, after a Hubbard-Stratonovich transformation,
\begin{widetext}
\begin{equation} \label{eq:E_int_Delta}
    E_{\bm q}^{\rm int}[\theta,\phi,\Delta] = \int d\bm r\, \frac{|\Delta_{\bm q}(\bm r)|^2}{U} - \frac{1}{2} \int d\bm r\,\left[ \Delta_{\bm q}(\bm r) \sum_{\bm k} e^{-i\phi_{\bm k}(\bm q)} \sin \theta_{\bm k} \, u^*_{\bm k+\bm q}(\bm r) u_{\bm k-\bm q}(\bm r) + c.c.\right].
\end{equation}
Here the complex field $\Delta$ is another variational parameter in addition to $\theta$ and $\phi$ to minimize the total energy. By the same argument that leads to Eq.~\eqref{eq:D_max_phi}, for the calculation of superfluid stiffness the $\bm q$-dependence of $\theta$ and $|\Delta|$ can be neglected. Furthermore, the second term in Eq.~\eqref{eq:E_int_Delta} is minimized by choosing $\phi$ such that each term in the $\bm k$-summation is real and positive. The remaining variational parameter is $\delta \equiv \arg\Delta$ and the geometric part of superfluid stiffness can be formally expressed as
\begin{equation}
    D_{ij}^{\rm geom} = - \frac{1}{A} \frac{\partial^2}{\partial q_i \partial q_j} \max_{\delta} \left\{\sum_{\bm k} \sin \theta_{\bm k} \left|\int d\bm r\, \Delta_0(\bm r) e^{i\delta_{\bm q}(\bm r)} u^*_{\bm k+\bm q}(\bm r) u_{\bm k-\bm q}(\bm r) \right|\right\}.
\end{equation}
\end{widetext}
For $\bm q\to 0$, the above expression is clearly a sum of local geometric quantities in momentum space. If $\Delta_0(\bm r)$ is a position-independent constant (uniform pairing condition \footnote{If $u_{\bm k}(\bm r)$ is nonzero only in a few isolated regions within a unit cell, as is the case for a tightly bound lattice model, the uniform pairing condition can be relaxed to requiring that $\Delta_0$ is constant within the concentration regions of $u_{\bm k}(\bm r)$.}), $D_{ij}^{\rm geom}$ involves a metric-like quantity
\begin{equation}
    g_{\bm k, ij}^{\delta} \equiv -\frac 14 \frac{\partial^2}{\partial q_i \partial q_j} \left|\int d\bm r\, e^{i\delta_{\bm q}(\bm r)} u^*_{\bm k+\bm q}(\bm r) u_{\bm k-\bm q}(\bm r) \right|
\end{equation}
that is generalized to include an intra-unit-cell phase twist $\delta$. For a flat-band system, $\theta_{\bm k}$ is a $\bm k$-independent constant and $D_{ii}^{\rm geom}$ is proportional to the minimal (integrated) quantum metric for all possible $\delta$:
\begin{equation}
    D_{ii}^{\rm geom} \propto \sum_{\bm k} g_{\bm k,ii}^{\rm min} \equiv \min_{\delta} \sum_{\bm k} g_{\bm k,ii}^{\delta}.
\end{equation}
For a continuum model, the physical quantum metric $g_{\bm k} = g_{\bm k}^{\delta=0}$ is well defined without any embedding ambiguity and provides an upper bound for superfluid stiffness. The physical quantum metric and the minimal quantum metric are in general different and coincide only in the presence of symmetry constraints. For example, when time-reversal and inversion symmetries are both preserved, $u_{\bm k}(\bm r) \in \mathbb{R}$ (up to a gauge choice) so $\delta=0$ minimizes the quantum metric.

In the lattice models considered in Refs. ~\cite{peotta2015superfluidity, julku2016geometric, liang2017band, chen2024ginzburg, huhtinen2022revisiting, peotta2025quantum, tam2024geometry, jiang2024geometric, herzog2022superfluid, herzog2022many}, $\bm r$ is replaced by a discrete sublattice label $\alpha$, the phase $\delta_{\bm q}^{\alpha} \approx \bm q \cdot \partial_{\bm q} \delta_{\bm q}^{\alpha}$ can be absorbed into the real-space position of sublattices \cite{huhtinen2022revisiting, peotta2025quantum, tam2024geometry}, so $g_{\bm k}^{\rm min}$ is equivalently defined as the minimal quantum metric for all possible sublattice embeddings.  

{\it Discussion.---}
The calculation of mean-field superfluid stiffness in general requires self-consistent calculations at multiple pairing momenta $\bm q$. Recent theoretical progress \cite{peotta2015superfluidity, julku2016geometric, liang2017band, torma2022superconductivity, xie2020topology, hu2019geometric, julku2020superfluid, chen2024ginzburg, huhtinen2022revisiting, peotta2025quantum, tam2024geometry, jiang2024geometric, herzog2022superfluid, herzog2022many} allows one to obtain superfluid stiffness based on knowledge of the $\bm q=0$ superconducting state only and relates the superflulid stiffness of flat-band superconductors to the quantum metric of single-particle Bloch states. This elegant result, however, requires restrictive assumptions that do not apply to recently studied two-dimensional materials. In particular, it requires that the effective attractive interaction is a short-range contact interaction that acts within each flavor of electrons. While this is true for Hubbard-type models (Eq.~\ref{eq:V_proj_Hubbard}) in which the relevant degrees of freedom for quantum geometry are real-space sublattices, it does not apply to graphene systems with microscopic layer and sublattice degrees of freedom. The projected interaction in the general case (Eqs.~\ref{eq:V_proj_cont},~\ref{eq:V_proj_moire}) mixes these degrees of freedom and as a result, local geometric quantities are not directly relevant.

It is important to emphasize that nontrivial quantum geometry by itself does not necessarily lead to an extra contribution to superfluid stiffness. In fact, we showed that it is possible to construct a model (continuum band with form factor \eqref{eq:Landau_form_factor}) in which the nontrivial quantum geometry does not break the effective Galilean invariance and therefore does not contribute to superfluid stiffness. Our theory suggests that the breaking of Galilean invariance in the interaction term should serve as the criterion of quantum geometric contribution to superfluid stiffness. The lower bound \eqref{eq:Ds_lower} can serve as an estimate of $D^{\rm geom}$ without performing self-consistent calculations at multiple $\bm q$'s. Our theoretical construction is reminiscent of our previous work \cite{zeng2024berry} on sliding electron crystals from a parent band with nontrivial quantum geometry and can be similarly applied to investigate quantum geometry effects on the Drude weight of Fermi liquids.

Recent studies of superconductivity in rhombohedral multilayer graphene \cite{zhou2021superconductivity, zhou2022isospin, zhang2023enhanced, han2024signatures} have raised new questions about quantum geometry effects on superconductivity in continuum bands. Our phenomenological calculations in Fig.~\ref{fig:graphene} suggest that despite the flat band bottom, the quantum geometric contribution to superfluid stiffness in these systems is less significant than in isolated flat bands. 
It remains to be explored how quantum geometry affects other aspects of superconductivity and, combined with the enhanced superfluid stiffness, how to utilize band geometry to engineer superconductors with increasing transition temperatures.

{\it Acknowledgments.---}
We thank Nishchhal Verma, Nicol\'as Morales-Dur\'an, Daniele Guerci, and Jonah Herzog-Arbeitman for helpful discussions. We thank P\"aivi T\"orm\"a for comments on a previous version of the manuscript.
Y.Z. and A.J.M. acknowledge support from Programmable Quantum Materials, an Energy Frontiers Research Center funded by the U.S. Department of Energy (DOE), Office of Science, Basic Energy Sciences (BES), under award DE-SC0019443.
The Flatiron Institute is a division of the Simons Foundation.

\bibliography{references}


\onecolumngrid
\newpage
\makeatletter 

\begin{center}
\textbf{\large Supplemental material for ``\@title ''} \\[10pt]
\end{center}
\vspace{20pt}

\setcounter{figure}{0}
\setcounter{section}{0}
\setcounter{equation}{0}

\renewcommand{\thefigure}{S\@arabic\c@figure}
\renewcommand{\theequation}{S\@arabic\c@equation}
\makeatother



\section{Derivation of mean-field superfluid stiffness}
Consider a system of spinful electrons in an isolated band with dispersion $\epsilon_{\bm k \uparrow} = \epsilon_{-\bm k \downarrow} \equiv \epsilon_{\bm k}$. With attractive interactions between opposite spins, the projected Hamiltonian reads
\begin{equation}
    H = \sum_{\bm k \sigma} \epsilon_{\bm k \sigma} c_{\bm k \sigma}^{\dagger} c_{\bm k \sigma} - \frac 1A \sum_{\bm k \bm k' \bm q} V_{\bm{kk}'}(\bm q) c_{\bm q+\bm k \uparrow}^{\dagger} c_{\bm q-\bm k \downarrow}^{\dagger} c_{\bm q-\bm k' \downarrow} c_{\bm q+\bm k' \uparrow}.
\end{equation}
Here $A$ is the size of the system and $V_{\bm{kk}'}(\bm q)$ is the electron-electron interactions projected onto the isolated band and encodes information about quantum geometry of the band; its explicit expression in terms of form factors will be presented below. Because of time-reversal symmetry, $V_{\bm{kk}'}(-\bm q) = V_{\bm{kk}'}^*(\bm q)$. In particular, $V_{\bm{kk}'}(\bm q=0) \in \mathbb{R}$ and $\partial_{\bm q} V_{\bm{kk}'}(\bm q) |_{\bm q=0}$ is purely imaginary.

Assuming spin-singlet pairing, a general BCS state with pairing momentum $\bm q$ takes the form
\begin{equation}
    \ket{\Phi_{\bm q}} = \prod_{\bm k} \left[\cos\frac{\theta_{\bm k}(\bm q)}{2} + e^{i\phi_{\bm k}(\bm q)} \sin\frac{\theta_{\bm k}(\bm q)}{2} c_{\bm q+\bm k \uparrow}^{\dagger} c_{\bm q-\bm k \downarrow}^{\dagger} \right] \ket{0},
\end{equation}
where $\ket{0}$ is the vacuum state (i.e. empty band). The total free energy (grand potential) of the above state is
\begin{align} \label{eq:Omega_q}
    \Omega_{\bm q}[\theta,\phi] = \braket{\Phi_{\bm q} | H-\mu N | \Phi_{\bm q}} = &\sum_{\bm k} (\epsilon_{\bm k + \bm q} + \epsilon_{\bm k - \bm q} - 2\mu) \sin^2 \frac{\theta_{\bm k}(\bm q)}{2} \nonumber\\
    &- \frac{1}{4A} \sum_{\bm k \bm k'} V_{\bm{kk}'}(\bm q) e^{i[\phi_{\bm k'}(\bm q) - \phi_{\bm k}(\bm q)]} \sin \theta_{\bm k}(\bm q) \sin \theta_{\bm k'}(\bm q).
\end{align}
For each $\bm q$, the self-consistent BCS state is obtained by minimizing $\Omega_{\bm q}$ in $(\theta,\phi)$ space, i.e.,
\begin{align}
    \frac{\partial}{\partial\theta_{\bm k}} \Omega_{\bm q} &= \frac 12 (\epsilon_{\bm k + \bm q} + \epsilon_{\bm k - \bm q} - 2\mu) \sin \theta_{\bm k}(\bm q) - \frac{1}{2A} \cos \theta_{\bm k}(\bm q) \sum_{\bm k'} {\rm Re}\{V_{\bm{kk}'}(\bm q) e^{i[\phi_{\bm k'}(\bm q) - \phi_{\bm k}(\bm q)]}\} \sin \theta_{\bm k'}(\bm q) = 0, \label{eq:MF_theta} \\
    \frac{\partial}{\partial\phi_{\bm k}} \Omega_{\bm q} &= -\frac{1}{2A} \sin \theta_{\bm k}(\bm q) \sum_{\bm k'} {\rm Im}\{V_{\bm{kk}'}(\bm q)  e^{i[\phi_{\bm k'}(\bm q) - \phi_{\bm k}(\bm q)]}\} \sin \theta_{\bm k'}(\bm q) = 0. \label{eq:MF_phi}
\end{align}
The above mean-field equations are equivalently derived from the Bogoliubov-de Gennes (BdG) Hamiltonian
\begin{equation}
    \mathcal{H}_{\rm BdG}(\bm q) = \sum_{\bm k} \begin{pmatrix}
        c_{\bm q + \bm k \uparrow}^{\dagger} & c_{\bm q - \bm k \downarrow}
    \end{pmatrix}
    \begin{pmatrix}
        \epsilon_{\bm k + \bm q} - \mu & -\Delta_{\bm k}(\bm q) \\
        -\Delta_{\bm k}^*(\bm q) & \mu - \epsilon_{\bm k - \bm q}
    \end{pmatrix}
    \begin{pmatrix}
        c_{\bm q + \bm k \uparrow} \\
        c_{\bm q - \bm k \downarrow}^{\dagger}
    \end{pmatrix}
\end{equation}
whose lower eigenvector is $(e^{i\phi_{\bm k}(\bm q)} \sin(\theta_{\bm k}(\bm q)/2), \cos(\theta_{\bm k}(\bm q)/2))^T$, with self-consistency condition
\begin{equation}
    \Delta_{\bm k}(\bm q) = \frac 1A \sum_{\bm k'} V_{\bm{kk}'}(\bm q) \braket{c_{\bm q - \bm k' \downarrow} c_{\bm q + \bm k' \uparrow}} = \frac{1}{2A} \sum_{\bm k'} V_{\bm{kk}'}(\bm q) e^{i\phi_{\bm k'}(\bm q)} \sin\theta_{\bm k'}(\bm q).
\end{equation}
At $\bm q=0$, because $V_{\bm{kk}'}(0)\in\mathbb{R}$, the solution is real: $\phi_{\bm k}(0) = 0$. Given the solution $\theta_{\bm k}$ at $\bm q=0$, our goal is to derive an expression of the superfluid stiffness
\begin{equation}
    [D_s]_{ij} = \frac{1}{A} \frac{d^2 \Omega_{\bm q}}{dq_i dq_j}\Bigg|_{\bm q=0}.
\end{equation}
Following the notation of Ref.~\cite{huhtinen2022revisiting}, we denote by $d/dq_i$ the derivative with $q_i$ that takes into account the change of $(\theta_{\bm k}, \phi_{\bm k})$ with $\bm q$. Calculating $D_s$ therefore requires knowledge of $(\theta_{\bm k}, \phi_{\bm k})$ at small nonzero $\bm q$ which is obtained by taking derivatives of the mean-field equations \eqref{eq:MF_theta}-\eqref{eq:MF_phi}:
\begin{equation} \label{eq:MF_dq}
    \frac{d}{d\bm q} \frac{\partial\Omega_{\bm q}}{\partial\theta_{\bm k}} \Bigg|_{\bm q=0} = 0, \quad \frac{d}{d\bm q} \frac{\partial\Omega_{\bm q}}{\partial\phi_{\bm k}} \Bigg|_{\bm q=0} = 0.
\end{equation}
The first equation involves only the first-order $\bm q$-derivative of $\theta_{\bm k}(\bm q)$ and its vanishing implies $\partial_{\bm q} \theta_{\bm k}(\bm q)|_{\bm q=0} = 0$. The second equation provides a condition for $\partial_{\bm q}\phi_{\bm k}$:
\begin{equation}
    \sum_{\bm k'} \left[{\rm Im}(\partial_{\bm q} V_{\bm{kk}'}) + |V_{\bm{kk}'}| \partial_{\bm q} (\phi_{\bm k'}-\phi_{\bm k}) \right] \sin\theta_{\bm k'} = 0.
\end{equation}
Here and below we set $\bm q=0$ in all expressions after taking derivatives. The explicit expressions of $\partial_{\bm q}\phi_{\bm k}$ can in principle be obtained formally by a matrix inversion. The superfluid stiffness can be written as
\begin{equation} \label{eq:Ds_dphi}
    [D_s]_{ij} = \frac{1}{A} \frac{d^2 \Omega}{dq_i dq_j} = \frac{1}{A} \frac{d}{dq_i} \frac{\partial\Omega}{\partial q_j} = \frac{1}{A} \frac{\partial^2 \Omega}{\partial q_i \partial q_j} + \frac{1}{A} \sum_{\bm k} \frac{\partial^2 \Omega}{\partial\phi_{\bm k} \partial q_j} \frac{\partial\phi_{\bm k}}{\partial q_i} = \frac{1}{A} \frac{\partial^2 \Omega}{\partial q_i \partial q_j} - \frac{1}{A} \sum_{\bm{kk}'} \frac{\partial^2 \Omega}{\partial\phi_{\bm k} \partial\phi_{\bm k'}} \frac{\partial\phi_{\bm k}}{\partial q_i} \frac{\partial\phi_{\bm k'}}{\partial q_j},
\end{equation}
where we made use of Eqs.~\eqref{eq:MF_theta}, \eqref{eq:MF_phi}, and \eqref{eq:MF_dq} and the vanishing of $\partial_{\bm q}\theta_{\bm k}$. Since the $\bm q$-dependence of $\theta_{\bm k}$ is unimportant, from Eq.~\eqref{eq:Omega_q} we take derivatives with $\bm q$ and get
\begin{equation}
    [D_s]_{ij} = \frac{2}{A} \sum_{\bm k} \frac{\partial^2 \epsilon_{\bm k}}{\partial k_i\partial k_j} \sin^2 \frac{\theta_{\bm k}}{2} - \frac{1}{4A^2} \sum_{\bm k\bm k'} \frac{\partial^2}{\partial q_i \partial q_j} \left[V_{\bm{kk}'} e^{i(\phi_{\bm k'}-\phi_{\bm k})} \right] \sin \theta_{\bm k} \sin \theta_{\bm k'}.
\end{equation}
The first term is the conventional contribution to superfluid stiffness that depends only on dispersion and the second term is the geometric contribution. Below we focus on the diagonal components ($i=j$) of $D_s$ and assume (approximate) rotational invariance. While an exact expression is not available without the explicit expression of $\phi_{\bm k}(\bm q)$, a general lower bound for the superfluid stiffness is obtained by taking the absolute value of the complex summands in the second term:
\begin{equation} \label{eq:Ds_lower_SM}
    D_s \ge \frac{2}{A} \sum_{\bm k} \epsilon_{\bm k}'' \sin^2 \frac{\theta_{\bm k}}{2} - \frac{1}{4A^2} \sum_{\bm k\bm k'} \frac{\partial^2}{\partial q^2} \big|V_{\bm{kk}'}(\bm q)\big| \sin \theta_{\bm k} \sin \theta_{\bm k'}.
\end{equation}
On the other hand, any choice of $\phi_{\bm k}(\bm q)$ provides an upper bound for $\Omega_{\bm q}$ and therefore an upper bound for $D_s$.

\subsection{Continuum band}
Consider a system with continuous translation symmetry like rhombohedral graphene gapped by a displacement field. In such systems the momentum space is unbounded and the projected interaction takes the form
\begin{equation}
    V_{\bm{kk}'}(\bm q) = V_{\bm k - \bm k'} \Lambda_{\bm k + \bm q, \bm k' + \bm q} \Lambda_{\bm k' - \bm q, \bm k - \bm q},
\end{equation}
where $\Lambda_{\bm{kk}'} \equiv \braket{u_{\bm k}^{\uparrow} | u_{\bm k'}^{\uparrow}} = \braket{u_{-\bm k'}^{\downarrow} | u_{-\bm k}^{\downarrow}}$. A lower bound for superfluid stiffness is readily obtained:
\begin{equation}
    D_s \ge \frac{2}{A} \sum_{\bm k} \epsilon_{\bm k}'' \sin^2 \frac{\theta_{\bm k}}{2} - \frac{1}{4A^2} \sum_{\bm k \bm k'} V_{\bm k - \bm k'} \sin \theta_{\bm k} \sin \theta_{\bm k'} \frac{\partial^2}{\partial q^2} \big|\Lambda_{\bm k + \bm q, \bm k' + \bm q} \Lambda_{\bm k' - \bm q, \bm k - \bm q}\big|.
\end{equation}
To get a gauge-invariant upper bound, we take $\phi_{\bm k} = \arg\braket{u_{\bm k+\bm q} | u_{\bm k-\bm q}}$ and it follows that
\begin{equation}
    D_s \le \frac{2}{A} \sum_{\bm k} \epsilon_{\bm k}'' \sin^2 \frac{\theta_{\bm k}}{2} - \frac{1}{4A^2} \sum_{\bm k \bm k'} V_{\bm k - \bm k'} \sin \theta_{\bm k} \sin \theta_{\bm k'} \frac{\partial^2}{\partial q^2} \left[\big|\Lambda_{\bm k + \bm q, \bm k' + \bm q} \Lambda_{\bm k' - \bm q, \bm k - \bm q}\big| \cos\varphi_{\bm{kk'q}} \right],
\end{equation}
where
\begin{equation}
    \varphi_{\bm{kk'q}} \equiv \arg[\braket{u_{\bm k-\bm q} | u_{\bm k+\bm q}} \braket{u_{\bm k+\bm q} | u_{\bm k'+\bm q}} \braket{u_{\bm k'+\bm q} | u_{\bm k'-\bm q}} \braket{u_{\bm k'-\bm q} | u_{\bm k-\bm q}}]
\end{equation}
is the gauge-invariant Berry phase around the loop $\bm k-\bm q \to \bm k+\bm q \to \bm k'+\bm q \to \bm k'-\bm q \to \bm k-\bm q$.

\subsection{Minibands in periodic systems}
For periodic systems such as moir\'e materials, the momenta $\bm k,\bm k'$ live within a Brillouin zone with reciprocal lattice vectors $\bm g$. The interaction potential takes the general form
\begin{equation}
    V_{\bm{kk}'}(\bm q) = \sum_{\bm g} V_{\bm k-\bm k'+\bm g} \braket{u_{\bm k+\bm q} | e^{i\bm g\cdot\bm r} | u_{\bm k'+\bm q}} \braket{u_{\bm k'-\bm q} | e^{-i\bm g\cdot\bm r} | u_{\bm k-\bm q}}
\end{equation}
where $\ket{u_{\bm k}} \equiv \ket{u_{\bm k}^{\uparrow}} = \ket{u_{-\bm k}^{\downarrow}}^*$. A lower bound for $D_s$ follows from Eq.~\eqref{eq:Ds_lower_SM}. To connect with the results from previous work, we assume contact interaction $V_{\bm k-\bm k'} = U$ and the total energy takes the form
\begin{align}
    \Omega_{\bm q}[\theta,\phi] &= \sum_{\bm k} (\epsilon_{\bm k + \bm q} + \epsilon_{\bm k - \bm q} - 2\mu) \sin^2 \frac{\theta_{\bm k}}{2} \nonumber \\
    &- \frac{U}{4} \int d\bm r \sum_{\bm k} e^{-i\phi_{\bm k}(\bm q)} \sin \theta_{\bm k} \, u^*_{\bm k+\bm q}(\bm r) u_{\bm k-\bm q}(\bm r) \sum_{\bm k'} e^{i\phi_{\bm k'}(\bm q)} \sin \theta_{\bm k'} \, u^*_{\bm k'-\bm q}(\bm r) u_{\bm k'+\bm q}(\bm r).
\end{align}
Here we assume that microscopic internal degrees of freedom such as sublattice and layer in multilayer graphene are irrelevant; quantum geometry comes from the momentum dependence of real-space Bloch wave functions $u_{\bm k}(\bm r)$. By a Hubbard-Stratonovich transformation,
\begin{align}
    \Omega_{\bm q}[\theta,\phi,\Delta] &= \sum_{\bm k} (\epsilon_{\bm k + \bm q} + \epsilon_{\bm k - \bm q} - 2\mu) \sin^2 \frac{\theta_{\bm k}}{2} + \int d\bm r\, \frac{|\Delta_{\bm q}(\bm r)|^2}{U} \nonumber \\
    &- \frac{1}{2} \int d\bm r\,\left[ \Delta_{\bm q}(\bm r) \sum_{\bm k} e^{-i\phi_{\bm k}(\bm q)} \sin \theta_{\bm k} \, u^*_{\bm k+\bm q}(\bm r) u_{\bm k-\bm q}(\bm r) + c.c.\right],
\end{align}
the minimization problem now becomes the minimization of $\Omega_{\bm q}$ in $(\theta,\phi,\Delta)$ space. The superfluid stiffness $D_s$ is obtained by taking the second total derivative of $\Omega_{\bm q}$ with $\bm q$ that takes into account the change of $(\theta,\phi,\Delta)$ with $\bm q$. Because $\phi=0$ and $\Delta\in\mathbb{R}$ at $\bm q=0$, by the same argument that leads to Eq.~\eqref{eq:Ds_dphi}, the variation of $\theta$ and $|\Delta|$ with $\bm q$ are unimportant; the variational parameters are $\phi_{\bm k}(\bm q)$ and the phase of $\Delta_{\bm q}(\bm r)$. Minimizing $\Omega$ with $\phi$ is easy -- just to choose $\phi$ such that each summand in $\bm k$ is real:
\begin{equation}
    \phi_{\bm k}(\bm q) = \arg\left[\int d\bm r\, \Delta_{\bm q}(\bm r) u^*_{\bm k+\bm q}(\bm r) u_{\bm k-\bm q}(\bm r) \right].
\end{equation}
The only remaining variational parameters are the phase of $\Delta_{\bm q}(\bm r)$. The superfluid stiffness can be written as
\begin{equation}
    D_s = \frac{2}{A} \sum_{\bm k} \epsilon_{\bm k}'' \sin^2 \frac{\theta_{\bm k}}{2} - \frac{1}{A} \max_{\delta} \left\{\sum_{\bm k} \sin \theta_{\bm k}\, \frac{\partial^2}{\partial q^2} \left|\int d\bm r\, \Delta_0(\bm r) e^{i\delta_{\bm q}(\bm r)} u^*_{\bm k+\bm q}(\bm r) u_{\bm k-\bm q}(\bm r) \right|\right\},
\end{equation}
where $\delta_{\bm q}(\bm r)$ is the phase of $\Delta_{\bm q}(\bm r)$. If $\Delta_0$ is a position-independent constant (i.e. uniform pairing), from the mean-field equation \eqref{eq:MF_theta} we can show that $\sin \theta_{\bm k} = \Delta_0/E_{\bm k}$ where $E_{\bm k} = \sqrt{(\epsilon_{\bm k}-\mu)^2 + \Delta_0^2}$ is the energy of Bogoliubov quasiparticles. The geometric part of $D_s$ is
\begin{align}
    D_s^{\rm geom} &= -\frac{\Delta_0^2}{A} \max_{\delta} \left\{\sum_{\bm k} \frac{1}{E_{\bm k}} \frac{\partial^2}{\partial q^2} \left|\int d\bm r\, e^{i\delta_{\bm q}(\bm r)} u^*_{\bm k+\bm q}(\bm r) u_{\bm k-\bm q}(\bm r) \right|\right\} \nonumber \\
    &= -\Delta_0^2\, \max_{\delta} \left\{\int\frac{d^d \bm k}{(2\pi)^d} \frac{1}{E_{\bm k}} \frac{\partial^2}{\partial q^2} \left|\braket{u_{\bm k+\bm q} | e^{i\delta_{\bm q}} | u_{\bm k-\bm q}} \right|\right\}.
\end{align}
If, furthermore, the isolated band is flat, $E_{\bm k} = E_0$ is independent of $\bm k$ and
\begin{equation}
    D_s^{\rm geom} = -\frac{\Delta_0^2}{E_0}\, \max_{\delta} \left\{\int\frac{d^d \bm k}{(2\pi)^d} \frac{\partial^2}{\partial q^2} \left|\braket{u_{\bm k+\bm q} | e^{i\delta_{\bm q}} | u_{\bm k-\bm q}} \right|\right\} \equiv \frac{4\Delta_0^2}{E_0} \int \frac{d^d\bm k}{(2\pi)^d} g_{\bm k}^{\rm min}.
\end{equation}
Here $g_{\bm k}^{\rm min}$ is the continuum-model generalization of the ``minimal quantum metric" introduced in Ref.~\cite{huhtinen2022revisiting}. Lattice models are special cases of continuum models where $u_{\bm k}(\bm r)$ is small except near a few points (``sublattices") in a unit cell, and uniform pairing means that $\Delta(\bm r)$ is uniform near each sublattice and effectively equal for all sublattices. Because of the embedding degree of freedom in lattice models, the phase $\delta_{\bm q}(\bm r)$ at each sublattice can be absorbed into the real-space location of the sublattice. The minimal quantum metric in lattice models can therefore be defined as the minimal value of quantum metric for all possible sublattice embeddings.

\section{Superfluid stiffness in Landau levels}
In this section we consider superconductivity in Landau levels. We first consider a fictitious system of two lowest Landau levels in opposite magnetic fields labeled by $\uparrow$ and $\downarrow$. The Landau levels are dispersionless and the wave functions are, in the Landau gauge,
\begin{equation}
    \psi_{k\uparrow}(\bm r) = \psi_{-k\downarrow}(\bm r)^* = \frac{1}{(\pi^{1/2} lL_y)^{1/2}} e^{iky} e^{-(x-kl^2)^2/2l^2},
\end{equation}
where $l = 1/\sqrt{eB}$ is the magnetic length and $L_y$ is the width of the system in $y$-direction. The Landau-gauge wave functions are localized in $x$-direction at the guiding center $kl^2$ and delocalized along $y$ with momentum $k$. Assuming attractive interaction between electrons in opposite Landau levels, the projected interaction takes the form
\begin{equation}
    H_{\rm int} = -\frac{1}{A} \sum_{kk'q} V_{kk'}(q) c_{q+k \uparrow}^{\dagger} c_{q-k \downarrow}^{\dagger} c_{q-k' \downarrow} c_{q+k' \uparrow},
\end{equation}
with
\begin{equation}
    V_{kk'}(q) = \frac{L_x}{4\pi} \int dp\, V_{p,k-k'}\, e^{-[p^2+(k-k')^2]l^2/2} \cos(2pql^2),
\end{equation}
where $V_{p,k-k'} \equiv V_{p\hat{\bm x} + (k-k')\hat{\bm y}}$ is the unprojected interaction in momentum space. Because the projected interaction is real, the lower bound \eqref{eq:Ds_lower} saturates and the mean-field superfluid stiffness has the exact expression
\begin{equation}
    D = -\frac{1}{4A^2} \sum_{kk'} \frac{\partial^2}{\partial q^2} \big|V_{kk'}(q)\big|_{q=0} \sin\theta_k \sin\theta_{k'} = \frac{l^2}{4\pi} \sin^2 \theta \int\frac{dpdk}{(2\pi)^2}\, p^2 V_{p,k} e^{-(p^2+k^2)l^2/2},
\end{equation}
where we have used the fact that the mean-field ground state for the lowest Landau level has $k$-independent constant $\theta_k=\theta$ and converted the $k$-summation to an integral $\sum_k \to L_y \int dk/2\pi$. The last expression can be simplified using rotation symmetry:
\begin{equation}
    D = \frac{l^2}{8\pi} \sin^2 \theta \int\frac{dpdk}{(2\pi)^2}\, (p^2+k^2) V_{p,k} e^{-(p^2+k^2)l^2/2} = \frac{l^2}{8\pi} \sin^2 \theta \int\frac{dp}{2\pi}\, p^3 V_{p} e^{-p^2 l^2/2}.
\end{equation}
When the Landau levels are half-filled, $\sin^2\theta = 1/2$ and we reproduce the expression \cite{moon1995spontaneous} of the superfluid stiffness of exciton condensate in quantum Hall bilayers which is equivalent to our problem by a particle-hole transformation.

Next we consider a single-component lowest Landau level. The projected interaction can be similarly written in the form
\begin{equation}
    H_{\rm int} = -\frac{1}{A} \sum_{kk'q} V_{kk'}(q) c_{q+k}^{\dagger} c_{q-k}^{\dagger} c_{q-k'} c_{q+k'},
\end{equation}
but without the spin indices (assuming all $\uparrow$). Because the momentum boost $q$ simply shifts all Bloch states by $ql^2$ in real space: $\psi_{k+q}(\bm r) = e^{iqy} \psi_k(\bm r - ql^2 \hat{\bm x})$, it is easy to see that $V_{kk'}(q)$ is independent of $q$. As a result, the superfluid stiffness vanishes despite the nontrivial quantum geometry. This is an exact result beyond mean-field theory.

\section{Models for numerical examples}
\subsection{Rhombohedral tetralayer graphene}
The single-particle Hamiltonian of rhombohedral tetralayer graphene is an $8\times 8$ matrix in layer-sublattice space. In layer space,
\begin{equation} \label{eq:H_PLG}
H_{\mathrm{4LG}} = \begin{pmatrix}
h_0+3u_D/2 & h_1 & h_2 & 0 \\
h_1^{\dagger} & h_0+u_D/2 & h_1 & h_2 \\
h_2^{\dagger} & h_1^{\dagger} & h_0-u_D/2 & h_1 \\
0 & h_2^{\dagger} & h_1^{\dagger} & h_0-3u_D/2
\end{pmatrix},
\end{equation}
where each $h$ is a $2\times 2$ matrix in sublattice space. The diagonal blocks
\begin{equation}
h_0(\bm p) = \begin{pmatrix}
0 & v_0 p_- \\
v_0 p_+ & 0
\end{pmatrix},
\end{equation}
where $p_{\pm} = \chi_v p_x \pm ip_y$ with $\chi_v=\pm$ depending on the valley, describe the low-energy physics of Dirac electrons in each graphene layer. The off-diagonal blocks
\begin{equation}
h_1(\bm p) = \begin{pmatrix}
v_4 p_- & v_3 p_+ \\
\gamma_1 & v_4 p_-
\end{pmatrix}, \quad
h_2 = \begin{pmatrix}
0 & \gamma_2/2 \\
0 & 0
\end{pmatrix}
\end{equation}
describe hopping between neighboring and next-neighboring layers respectively. The effect of an applied displacement field is modelled as a potential difference $u_D$ between neighboring layers. We use parameters from Ref.~\cite{zhou2021half} which are chosen to match the experimental data in rhombohedral trilayer graphene: $v_0 = \SI{1.0e6}{m/s}$, $\gamma_1 = \SI{380}{meV}$, $\gamma_2 = \SI{-15}{meV}$, $v_3 = \SI{-9.4e4}{m/s}$, $v_4 = \SI{-4.6e4}{m/s}$.

\subsection{Lieb lattice}
The Lieb lattice model has three sublattices with intra-unit-cell hopping (solid lines in Fig.~\ref{fig:lieb}) $1+\eta$ and inter-unit-cell hopping (dotted lines) $1-\eta$. In the sublattice (red, green, blue) basis, the momentum-space Hamiltonian is a $3\times 3$ matrix
\begin{equation}
    H_{\rm{Lieb}}(\bm k) = \begin{pmatrix}
        0 & 1+\eta+(1-\eta)e^{ik_y} & 0 \\
        1+\eta+(1-\eta)e^{-ik_y} & 0 & 1+\eta+(1-\eta)e^{-ik_x} \\
        0 & 1+\eta+(1-\eta)e^{ik_x} & 0
    \end{pmatrix},
\end{equation}
where the hopping amplitude is set as the energy scale and the lattice constant is the length scale. Because the matrix has zero determinant for all $\bm k$, a zero-energy flat band exists. The other two eigenvalues are $\pm 2[1+\eta^2+(1-\eta^2)(\cos k_x+\cos k_y)/2]^{1/2}$. The gap at $(k_x,k_y) = (\pi,\pi)$ closes when $\eta=0$. When $\eta\ne 0$, the flat band is isolated from the remote bands by a gap of size $2\sqrt{2}\eta$.

\end{document}